\documentclass[num-refs]{wiley-article}

\usepackage{graphicx}
\usepackage[space]{grffile}
\usepackage{latexsym}
\usepackage{textcomp}
\usepackage{longtable}
\usepackage{tabulary}
\usepackage{booktabs,array,multirow}
\usepackage{amsfonts,amsmath,amssymb}
\usepackage{natbib}
\usepackage{comment}
\usepackage{algorithm}
\usepackage{algpseudocode}

\usepackage{url}
\usepackage{hyperref}
\hypersetup{colorlinks=false,pdfborder={0 0 0}}
\usepackage{etoolbox}
\makeatother
\newif\iflatexml\latexmlfalse

\AtBeginDocument{\DeclareGraphicsExtensions{.pdf,.PDF,.eps,.EPS,.png,.PNG,.tif,.TIF,.jpg,.JPG,.jpeg,.JPEG}}

\usepackage[utf8]{inputenc}
\usepackage[english]{babel}

\usepackage{siunitx}

\iflatexml

\else

\paperfield{Biostatistics – Survival Analysis}

\corraddress{Theofanis Balanos, Department of Biostatistics and Health Data Science, Indiana University Indianapolis, IN, USA.}
\corremail{tbalanos@iu.edu}

\fundinginfo{Research reported in this publication was supported by the National Institute Of Allergy And Infectious Diseases (NIAID), Eunice Kennedy Shriver National Institute Of Child Health \& Human Development (NICHD), National Institute On Drug Abuse (NIDA), National Cancer Institute (NCI), and the National Institute of Mental Health (NIMH), National Institute of Diabetes and Digestive and Kidney Diseases (NIDDK) , Fogarty International Center (FIC), National Heart, Lung, and Blood Institute (NHLBI) , National Institute on Alcohol Abuse and Alchoholism (NIAAA), in accordance with the regulatory requirements of the National Institutes of Health under Award Number U01AI069911 East Africa IeDEA Consortium. The content is solely the responsibility of the authors and does not necessarily represent the official views of the National Institutes of Health.}
\fi

\papertype{Original Article}

\title{Semiparametric Regression for Misclassified Competing Risks Data}

\author[1]{\normalsize Theofanis Balanos}
\author[2]{\normalsize Constantin T. Yiannoutsos}
\author[1]{\normalsize Felix M. Pabon-Rodriguez}
\author[3]{\normalsize Hongmei Nan}
\author[1]{\normalsize Giorgos Bakoyannis}

\affil[1]{Department of Biostatistics and Health Data Science, Fairbanks School of Public Health and School of Medicine, Indiana University Indianapolis, Indianapolis, IN, USA}
\affil[2]{Department of Epidemiology and Biostatistics, CUNY Graduate School of Public Health and Health Policy, City University of New York, New York, NY, USA}
\affil[3]{Department of Epidemiology, Richard M. Fairbanks School of Public Health, Indiana University Indianapolis, IN, USA}

\runningauthor{Balanos \textit{et al.}}

\usepackage{placeins}

\begin{document}

\maketitle
\selectlanguage{english}

\begin{abstract}
The analysis of competing risks data is often complicated by misclassification of the cause of failure. This issue can lead to seriously biased estimates and invalid conclusions. One way to deal with such misclassification is to use a gold-standard cause of failure ascertainment procedure in a subset of the non-right-censored participants (internal validation sample) along with methods for missing data to deal with the missing gold-standard ascertainments. However, this approach can be costly and time-consuming and, therefore, cannot be implemented in many studies. In this work, we propose a semiparametric regression analysis methodology for the case where no internal validation sample exists. Our approach leverages estimates of the misclassification probabilities from an external validation study to adjust for misclassification in the study at hand. These probabilities are incorporated in a B-spline-based sieve pseudo-likelihood function, which is maximized to jointly estimate models for all event types. Using empirical process theory, we show that the proposed estimator is consistent. Extensive simulation experiments demonstrate that the method performs well with realistic sample sizes and provides substantially more efficient estimates compared to previously proposed approaches. The methodology is applied to competing risks data from a large HIV observational study in sub-Saharan Africa, where event type is misclassified due to significant death under-reporting.

\vspace{0.5cm}
\end{abstract}

\section{Introduction}\label{sec:intro_p1}

Misclassification in competing risks data occurs when event type, also known as cause of failure, is incorrectly recorded \citep{bakoyannis2020semiparametric}. This issue presents a significant challenge in the analysis of such data since it can lead to bias and incorrect conclusions about the relationship between risk factors and the causes of failure under study \citep{barron1977effects, bross1954misclassification, lyles2011validation, magder1997logistic, neuhaus1999bias}. Misclassification often arises in real-world data due to errors in data collection, inaccurate reporting, or limitations in outcome verification procedures \citep{edwards2013accounting}. 
A standard approach for addressing misclassification is to adjust estimators based on known or estimated misclassification probabilities \citep{lyles2011validation, tang2015binary}. 
A common method for obtaining misclassification probabilities is by utilizing internal validation, often referred to as double-sampling. This approach involves applying a gold-standard cause of failure ascertainment procedure to a subset of the study population \citep{greenland1988variance, tenenbein1970double}. The misclassification probabilities can then be estimated by comparing the observed, and potentially misclassified, outcomes with the true outcomes in this internal-validation subsample. However, internal validation is resource intensive and often impractical, despite its well-established accuracy, especially in large-scale epidemiological and clinical research \citep{spiegelman2001efficient}. Thus, when internal validation is not feasible, externally estimated misclassification probabilities provide a viable alternative \citep{spiegelman2001efficient}.

An example of the need for misclassification correction can be found in studies in sub-Saharan Africa \citep{yiannoutsos2008sampling, an2009need, geng2008sampling, monroy2024alcohol}, involving people living with HIV/AIDS receiving care at various health facilities that are part of the East Africa International epidemiology Databases to Evaluate AIDS/HIV (EA-IeDEA) regional consortium. As a part of these studies, it is crucial to understand the factors associated with the hazard of a gap in care (i.e., missing a scheduled appointment and then remaining out of care for at least 60 days) and death. Misclassification between these two causes of failure is particularly common in this setting, as many deaths are under-reported, and these deaths are erroneously classified as gaps in care \citep{bakoyannis2020semiparametric, egger2011correcting, brinkhof2010adjusting}. This misclassification can result in underestimation of mortality rates and overestimation of gaps in HIV care, which may skew an accurate assessment of care programs. Also, it can lead to biased effect estimates of predictors of gap in care and death. This bias can be mitigated through internal validation. This involves tracking in the community a subset of patients who are labeled as \textit{lost} (i.e., missing a scheduled clinic visit for at least 60 days) to determine their true vital status \citep{bakoyannis2020semiparametric, brinkhof2010adjusting}. Despite the accuracy of internal validation estimates, this method is not practical, particularly in resource-constrained settings such as those found in sub-Saharan Africa. Thus, external validation, where misclassification probabilities are estimated using a separate dataset in which both observed and true causes of failure are available, serves as the more practical option, avoiding the cost and logistical challenges of patient tracing required by internal validation \citep{mpofu2020pseudo}.

The use of externally estimated misclassification probabilities for adjusting for misclassification in semiparametric competing risks regression analysis remains largely unexplored \citep{ha2015semiparametric, mpofu2020pseudo}. To the best of our knowledge, this has only been addressed in the work by Ha and Tsodikov \cite{ha2015semiparametric}. These authors proposed both likelihood-based and estimating-equations-based methods. The likelihood-based methods impose a strong parametric assumption on the relationship between the baseline hazards of the different causes. Given that a violation of this assumption can lead to bias, these authors proposed a weighted martingale estimating-equations approach, hereafter referred to as the Ha and Tsodikov weighted martingale method (HT-WM), to estimate a fully semiparametric proportional cause-specific hazards model that avoids this parametric assumption. Nevertheless, estimating-equation-type methods are expected to achieve inferior statistical efficiency compared to likelihood-based methods. In addition, this approach does not account for the additional uncertainty in cases where misclassification probabilities have been externally estimated. Furthermore, the issue of violation of the \textit{transportability assumption}, under which the misclassification probabilities from the external setting are applicable to the main study, was not considered by the authors \cite{ha2015semiparametric}. Last but not least, there is no software to readily implement these approaches. 

In this work, we propose a likelihood-based methodology for the semiparametric proportional cause-specific hazards model with misclassified competing risks data. Our proposed methodology relies on known or externally estimated misclassification probabilities, which are incorporated into our B-spline-based sieve pseudo-likelihood function. Importantly, we do not impose restrictive parametric assumptions on the baseline cause-specific hazard functions. Moreover, we explicitly account for the additional uncertainty in the externally estimated misclassification probabilities via an appropriate bootstrap algorithm. Notably, this algorithm does not require access to the external validation data and just relies on the variance matrix of the externally estimated parameters. Furthermore, we propose a sensitivity analysis approach for evaluating the robustness of our estimates against violations of the transportability assumption. In addition, we provide an easy-to-use R function to readily implement the proposed methodology. Using empirical process theory, we show that our estimator is consistent. Extensive simulation experiments show that our estimator works well in finite samples and that it is substantially more statistically efficient compared to the HT-WM approach \cite{ha2015semiparametric}. Finally, our proposed method is applied to data from the EA-IeDEA regional consortium to evaluate risk factors for gap in care and death, while accounting for misclassification due to death under-reporting.

The rest of the paper is organized as follows. Notation, models, and the proposed B-spline-based sieve pseudo-likelihood approach that adjusts for misclassified causes of failure are presented in Section~\ref{sec:method_p1}. In Section~\ref{sec:properties_p1}, we establish consistency of the proposed estimator under a set of realistic regularity conditions. In Section~\ref{sec:sens_p1}, we present a sensitivity analysis to account for possible violations of the transportability assumption. Section~\ref{sec:sim_p1} presents extensive simulation studies evaluating the finite-sample performance of our method and comparing it with the HT–WM method. In Section~\ref{sec:dataapp_p1}, we apply the method to competing risks data from the EA-IeDEA regional consortium. Finally, Section~\ref{sec:disc_p1} concludes with a brief discussion.

\section{Methodology}\label{sec:method_p1}

\subsection{Notation}
Let \(T\) and \(U\) be the failure time and the right censoring time, respectively. Now, denote by \(X = \min(T,U)\) the observed failure time, and let \(\Delta = I(T \leq U)\) be the failure indicator. Additionally, let \(C \in \{1, \ldots, k\}\) denote the true cause of failure and \(C^{*} \in \{1, \ldots, k\}\) represent the observed, potentially misclassified, cause of failure among \(k\) competing risks, where \(k\) is finite. We assume that the observation interval is \([0, \tau]\), with \(\tau < \infty\). Finally, denote by \(Z\) a \(p\)-dimensional vector of covariates.

The basic identifiable quantities based on competing risks data are the cause-specific hazards and the cumulative incidence functions \citep{kalbfleisch2002statistical}. The cause-specific hazard for cause \( j \), given covariates \( Z = z \), is defined as:
\[
\lambda_j(t; z) = \lim_{h \downarrow 0} \frac{1}{h} P(t \leq T < t + h, C = j \mid T \geq t, Z = z), \quad j = 1, \ldots, k.
    \]
This function describes the instantaneous failure rate due to cause \( j \) at time \( t \), given that the individual has survived up to \( t \). The cumulative incidence function for the \( j \)th cause of failure is:
    \[
    F_j(t; z) = P(T \leq t, C = j \mid Z = z), \quad j = 1, \ldots, k.
    \]    
This function represents the probability of failure due to cause \( j \) by time \( t \), given covariates \( z \). The cumulative incidence function for cause $j$ can be expressed as function of the cause-specific hazards for all causes \citep{bakoyannis2012practical}:
    \[
    F_j(t; z) = \int_0^t \exp\left( - \sum_{j=1}^k \Lambda_j(s; z) \right) \lambda_j(s; z) ds, \quad j = 1, \ldots, k,
    \]
where \( \Lambda_j(t; z) = \int_0^t \lambda_j(s; z) ds \) is the cumulative hazard function for the \( j \)th cause of failure. A standard model for the cause-specific hazards is the semiparametric proportional hazards model \citep{bakoyannis2012practical, kalbfleisch2002statistical}:
\[
\lambda_j(t; z) = \lambda_{0,j}(t) \exp(\beta_j^T z), \quad j = 1, \ldots, k,
\]
where \( \lambda_{0,j}(t) \) is the \( j \)th unspecified baseline cause-specific hazard function and \( \beta_j \) is the vector of covariate effects specific to cause \( j \).

\subsection{Misclassification probabilities}\label{sec:misclass_prob_p1}

In a setting with misclassification, that is, when \( P(C^* \neq C) > 0 \), the classification probabilities are defined as follows:
\[
P(C^* = j \mid C = h, T = t, Z = z) \equiv \pi_{jh}^*(t, z; \gamma_0), \quad j, h = 1, \ldots, k,
\]
where \( \pi_{jh}^*(t, z; \gamma_0) \) denotes the probability of observing cause \( j \) when the true cause is \( h \), conditional on covariates \( Z=z\) and failure time \( T=t \). The parameter vector \( \gamma_0\) is a finite-dimensional parameter.

Classical methods for estimating the cause-specific hazards and the corresponding cumulative incidence functions assume that the observed cause of failure is correctly recorded, i.e., \(C^* = C\), and are therefore expected to yield biased estimates when misclassification is present \citep{mpofu2020pseudo}. Given the classification probabilities, the misclassified cause-specific hazard
\[
\lambda_j^*(t; z) = \lim_{h \downarrow 0} \frac{1}{h} P(t \leq T < t + h, C^* = j \mid T \geq t, Z = z), \quad j = 1, \ldots, k,
\]
can be expressed in terms of the cause-specific hazards for the true cause $C$:
\begin{equation}
\lambda_j^*(t; z) = \sum_{h=1}^k \pi_{jh}^*(t, z; \gamma_0) \lambda_{0,h}(t) \exp(\beta_h^T z), \quad j = 1, \ldots, k.
\label{eq:lambda_star_p1}
\end{equation}
The expression above shows that the observable misclassified cause-specific hazard for cause \(j\) is a mixture of the true hazards across all causes, weighted by the (mis)classification probabilities \(\pi_{jh}^*(t, z; \gamma_0)\). Given this, we will introduce this definition into the likelihood function so that it depends on the parameters of interest \( \lambda_j(t; z) \) and the externally estimable misclassification probabilities \citep{mpofu2020pseudo}.

\subsection{Estimation approach}\label{sec:est_app_p1}

For each individual $i$, where $i = 1, 2, \ldots, n$, the observed data are ${(\Delta^{*}_{i1},\ldots,\Delta^{*}_{ik}, X_i, Z_i)}$, where $\Delta_{ij}^*=\Delta_i\,I(C_i^*=j)$ is the misclassified indicator of the $j$th cause of failure for the $i$th individual. The likelihood of the observed (i.e., misclassified) competing risks data is, under the assumption of independent and non-informative right-censoring, proportional to:
\[
L_n(\theta) \propto \prod_{i=1}^n \left\{ \exp \left[ - \sum_{j=1}^k \Lambda_j^*(X_i; Z_i) \right] 
\prod_{j=1}^k \left[ \lambda_j^*(X_i; Z_i) \right]^{\Delta^{*}_{ij}} \right\}.
\]
In light of~\eqref{eq:lambda_star_p1}, the likelihood function of the observed data is expressed as a function of the cause-specific hazards of interest:
\[
L_n(\theta) \propto \prod_{i=1}^n \left\{ \exp \left[ - \sum_{j=1}^k \Lambda_j(X_i; Z_i, \theta_j) \right] 
\prod_{j=1}^k \left[ \sum_{h=1}^k \pi_{jh}^*(X_i, Z_i; \gamma_0) \lambda_h(X_i; Z_i, \theta_j) \right]^{\Delta^*_{ij}} \right\}.
\]
If the cause of failure ascertainment is informative with respect to the true cause \(C\), that is
\[
\pi_{jj}^*(t, z; \gamma_0) > 0.5 \quad \text{for all} \quad j=1,\ldots,k,
\]
for all $(t,z)$, which means that the probability of correct classification (i.e., $C^*=j$ when $C=j)$ is over 0.5, then the model is identifiable based on the observed data. Under the proportional cause-specific hazards model, the cumulative cause-specific hazards can be expressed as:
\[
\Lambda_j(t; z, \theta_j) = \Lambda_{0,j}(t) \exp(\beta_j^T z) = \exp[\phi_j(t) + \beta_j^T z], \quad j = 1, \ldots, k,
\]
where \( \theta_j = (\beta_j^T,\phi_j)^T \) and \( \phi_j \) is an unspecified, strictly increasing, and invertible function of time. The corresponding cause-specific hazard function is:
\[
\lambda_j(t; z, \theta_j) = \exp[\phi_j(t) + \beta_j^T z] \phi_j'(t),
\]
where \( \phi_j'(t) = d \phi_j(t)/dt \). Therefore, the log-likelihood function is:
\begin{align*}
\ell_n(\theta; \gamma_0) &= 
\sum_{i=1}^{n} \left(
    \sum_{j=1}^{k} \Delta^*_{ij} \log 
    \left\{
        \sum_{h=1}^{k} \exp\left[\phi_h(X_i) + \beta_h^T Z_i\right] 
        \phi_h'(X_i) \, \pi_{jh}^*(X_i, Z_i; \gamma_0)
    \right\} \right. - \left. 
    \sum_{j=1}^{k} \exp\left[\phi_j(X_i) + \beta_j^T Z_i\right]
    \right).
\end{align*}
In practice, the true parameter \( \gamma_0 \) governing the misclassification probabilities is unknown and must be estimated using an external validation dataset. A natural choice for modeling the misclassification probabilities is the multinomial logit model with a generalized logit link function for \( k > 2 \), or the binary logit model with the logit link function when \( k = 2 \). For the special case of the binary logit model, the misclassification probability is:
\[
\pi_{jh}^*(X_i, Z_i; \gamma_0) = \frac{\exp\left(\gamma_0^T (1, X_i, Z_i^T)^T\right)}{1 + \exp\left(\gamma_0^T (1, X_i, Z_i^T)^T\right)},
\]
where \( (1, X_i, Z_i^T)^T \) represents the covariate vector for individual \( i \), including an intercept. In our motivating example, \( \pi_{12}^*(X_i, Z_i; \gamma_0) \) represents the probability of classifying a patient as having a gap in care ($C^*=1$) when they are actually deceased ($C=2$). In general, these probabilities are estimated using an external validation dataset of sample size $n'$, where the observed cause of failure is known for all individuals, but the true cause of failure (based on a gold-standard ascertainment procedure) is known only for a subset. In such settings, estimation of the misclassifiaction probabilities can be achieved using the approach proposed by Mpofu et al. \citep{mpofu2020pseudo}, to account for the missing gold standard cause of failure ascertainments. Thus, letting $\hat{\gamma}_{n'}$ denote an estimate of $\gamma_0$, the proposed log-pseudolikelihood function is:
\begin{equation}
\ell_n(\theta;\hat{\gamma}_{n'}) =
\sum_{i=1}^{n} \left(
    \sum_{j=1}^{k} \Delta^*_{ij} \log 
    \left\{
        \sum_{h=1}^{k} \exp\left[\phi_h(X_i) + \beta_h^T Z_i\right] 
        \phi_h'(X_i) \, \pi_{jh}^*(X_i, Z_i; \hat{\gamma}_{n'})
    \right\}
    - \sum_{j=1}^{k} \exp\left[\phi_j(X_i) + \beta_j^T Z_i\right]
\right).
\label{eq:logPL_p1}
\end{equation}
The unknown functions $\phi_{n,j}(t)$ and $\phi_{n,j}'(t)$ can be approximated by B-spline functions. Formally, define the space of monotone B-spline functions
\[
\Phi_{n,j}=\left\{ \phi_{n,j}'(t) = \sum_{s=1}^{N_{n,j} + m_j} \gamma_{j,s} B_{s,m_j}(t): \,\gamma_{j,1},\ldots, \gamma_{j,N_{n,j} + m_j}\in\mathbb{R},\,\gamma_{j,1}<\cdots< \gamma_{j,N_{n,j} + m_j},\, t \in [0, \tau] \right\},
\]
where $N_{n,j}$ denotes the number of internal knots, $m_j$ is the order of the spline, and \( B_{s,m_j}(t) \) are B-spline basis functions. Note that the number of internal knots increases with the sample size $n$, so that the approximation of the true functions $\phi_j$, $j=1,\ldots,k$, improves as $n\rightarrow\infty$ \cite{zhang2010spline,bakoyannis2017semiparametric}. It has been shown that the number of internal knots leading the optimal rate of convergence for the infinite-dimensional parameters $\phi_j$ is $N_{n,j}=O(n^{\nu})$, with $\nu=1/(1+2p)$, where $p$ is the degree of smoothness of the true $\phi_j$, $j=1,\ldots,k$ \cite{zhang2010spline,bakoyannis2017semiparametric}. The inequality in the control points of the spline ensures monotonicity, so that the estimated cumulative baseline cause-specific hazard functions are monotonic functions of time. Then $\phi_{n,j}\in\Phi_{n,j}$ and $\phi_{n,j}'$ is the first derivative of $\phi_{n,j}$. The proposed B-spline-based sieve maximum likelihood estimate $\hat{\theta}_n$ of $\theta_0$ is obtained by maximizing Equation~\eqref{eq:logPL_p1}. For the situation with known misclassification parameters $\gamma_0$, variance estimation can be performed using standard nonparametric bootstrap \cite{cheng2010bootstrap}. When misclassification parameters are not assumed to be known, variance estimation can be performed based on the estimated variance matrix $\hat{\Omega}_{n'}$ of $\hat{\gamma}_{n'}$ and the bootstrap-based Algorithm~\ref{alg:var}. A key feature of this algoritm is that it does not require access to the orignal external validation data. It only relies on the estimates  $\hat{\gamma}_{n'}$ and $\hat{\Omega}_{n'}$ from the external study. 

\begin{algorithm}[H]
\caption{Variance estimation with externally estimated misclassification probabilities}
\label{alg:var}
\begin{algorithmic}[1]
\Require Main study data $\{(\Delta_{i1}^{*},\ldots,\Delta_{ik}^{*},X_i, Z_i)\}_{i=1}^n$, misclassification parameter estimate $\hat{\gamma}_{n'}$, variance matrix $\hat{\Omega}_{n'}$ of $\hat{\gamma}_{n'}$, number of bootstrap replications $B$
\State Set bootstrap iteration counter $b \gets 0$
\Repeat
    \State $b \gets b + 1$
    \State Generate a simulation realization $\tilde{\gamma}_{n'}^{(b)}\sim N(\hat{\gamma}_{n'},\hat{\Omega}_{n'})$ of the misclassification parameter
    \State Draw a sample $\{(\Delta_{i1}^{*(b)},\ldots,\Delta_{ik}^{*(b)},X_i^{(b)}, Z_i^{(b)})\}_{i=1}^n$ from the data in the main study with replacement
    \State Using the sample $\{(\Delta_{i1}^{*(b)},\ldots,\Delta_{ik}^{*(b)},X_i^{(b)}, Z_i^{(b)})\}_{i=1}^n$ maximize $\ell_n(\theta;\tilde{\gamma}_{n'}^{(b)})$  to get $\hat{\theta}_n^{(b)}=(\hat{\beta}_n^{(b)},\hat{\phi}_n^{(b)})$
\Until{$b\leq B$}
\State Compute the variance matrix $\hat{\Sigma}_n$ of $\hat{\beta}_n$ as the empirical variance matrix of the sample $\hat{\beta}_n^{(1)},\ldots,\hat{\beta}_n^{(B)}$
\State \Return $\hat{\Sigma}_n$ 
\end{algorithmic}
\end{algorithm}

\section{Consistency of the proposed estimator}\label{sec:properties_p1}

Define the sieve parameter space
\[
\Theta_n=\times_{j=1}^{k}\big(\mathcal{B}_j \times \Phi_{n,j}\big),
\]
where the finite-dimensional parameters are $\beta_j\in\mathcal{B}_j$ and the infinite-dimensional parameters are
$\phi_j\in\Phi_{n,j}$, $j=1,\ldots,k$. Consistency is shown under the $L_2$-metric $d(\cdot,\cdot)$ defined by
\[
d\!\big(\theta^{(1)},\theta^{(2)}\big)
=\Bigg(\sum_{j=1}^{k}\big\|\beta^{(1)}_{j}-\beta^{(2)}_{j}\big\|^{2}
\;+\;\sum_{j=1}^{k}\big\|\phi^{(1)}_{j}-\phi^{(2)}_{j}\big\|_{\Phi}^{2}\Bigg)^{1/2},
\]
for $\theta^{(1)},\theta^{(2)}\in\Theta$, where $\|\cdot\|$ denotes the Euclidean norm and the norm for the infinite-dimensional parameters is the $L_2$ norm
\[
\big\|\phi^{(1)}_{j}-\phi^{(2)}_{j}\big\|_{\Phi}^{2}
=E\!\left[\big\{\phi^{(1)}_{j}(X)-\phi^{(2)}_{j}(X)\big\}^{2}\right],\qquad j=1,\ldots,k. 
\]
Finally, let $\theta_0$ denote the true parameter value. Theorem~\ref{thm:consistency_p1} establishes the consistency of the proposed estimator.

\begin{theorem}\label{thm:consistency_p1}
Suppose right censoring is independent of $(T,C)$ and noninformative conditionally on $Z$, the regularity conditions C1–C4 listed in the Appendix hold, the misclassification model $\pi^*_{jh}(T,Z;\gamma_0)$ is correctly specified with 
$\hat{\gamma}_{n'} \xrightarrow{p}\gamma_0$, and the number of interior knots 
$N_{n,j} = O(n^\nu)$ with $1/[2(1+p)] < \nu < 1/(2p)$ for $j=1,\ldots,k$. 
Then
\[
d\!\left(\hat{\theta}_{n},\theta_{0}\right)\xrightarrow{p}0,
\] 
as $\min(n,n')\rightarrow\infty$, with $n$ and $n'$ having the same growth rate.
\end{theorem}

The proof of Theorem~\ref{thm:consistency_p1} using empirical process theory techniques is outlined in the Appendix. Using some additional weak regularity conditions, together with arguments similar to those in Bakoyannis et al. \citep{bakoyannis2017semiparametric},
and noting that the external-validation estimator $\hat{\gamma}_{n'}$ is $\sqrt{n'}$-consistent, it can be shown that $\hat{\beta}_n$ is $\sqrt{n}$–consistent and asymptotically normal. Variance estimation can be performed using the bootstrap-based Algorithm~\ref{alg:var}. The use of nonparametric bootstrap for estimating the variance of general M-estimators of finite-dimensional parameters has previously been justified \citep{cheng2010bootstrap}. The simulation step in Algorithm~\ref{alg:var} is justified by the asymptotic normality of $\sqrt{n'}(\hat{\gamma}_{n'}-\gamma_0)$, and the consistency of $\hat{\gamma}_{n'}$ and its variance estimator $\hat{\Omega}_{n'}$, by virtue of this estimator being a parametric maximum likelihood estimator. This simulation step to account for the variability in $\hat{\gamma}_{n'}$ is similar to Rubin's type A multiple imputation \cite{rubin1996multiple, robins2000inference}.

\section{Sensitivity analysis}\label{sec:sens_p1}

Our proposed methodology is assumed to rely on the transportability assumption, that is, the misclassification probabilities from the external validation dataset are the same as those in the main study population. In this work, we address any potential violations of this assumption through sensitivity analysis. Specifically, we introduce a sensitivity parameter \( \eta \) that quantifies the discrepancy between the misclassification probabilities from the main and external validation studies. Let \(S = 1\) denote the main study sample and \(S = 0\) the external validation study sample. 
The transportability assumption is:
\begin{eqnarray*}
P(C^* = j \mid C = h, S = 0, T = t, Z = z) 
&=& P(C^* = j \mid C = h, S = 1, T = t, Z = z) \\
&=& g_{jh}\!\left(\gamma_{0,jh}^\top \tilde{W}\right),
\end{eqnarray*}
where \( g_{jh}(\cdot) \) denotes the inverse of a link function corresponding to the
misclassification probability from true cause \(h\) to observed cause \(j\),
\( \gamma_{0,jh} \) is the unknown parameter vector estimated using the external
validation sample, and \( \tilde{W} = (1, T, Z^T)^T \). However, if the transportability assumption is violated, then the last equality does not hold.
For such situations, we define the probability of misclassification as a function of the
sensitivity parameter \( \eta \):
\[
P(C^* = j \mid C = h, S, T = t, Z = z)
=
g_{jh}\!\left(\gamma_{0,jh}^\top \tilde{W} + \eta S\right).
\] If \( \eta=0 \), then transportability assumption is satisfied, while \( \eta \neq 0 \) implies a violation of this assumption. For the special case with $k=2$ and a logistic model for the probability of misclassification, the sensitivity parameter \( \eta \) is interpreted as a log odds ratio, representing the relative odds of misclassification (i.e., observing \( C^* = j \)) in the main dataset (\( S = 1 \)) compared to the external validation dataset (\( S = 0 \)), conditional on the true cause \( C = h \), event time \( T \), and covariates \( Z \):
\[
\eta = 
\log \left(
\frac{
P(C^* = j \mid C = h, S = 1, T, Z) \big/ \left[1 - P(C^* = j \mid C = h, S = 1, T, Z)\right]
}{
P(C^* = j \mid C = h, S = 0, T, Z) \big/ \left[1 - P(C^* = j \mid C = h, S = 0, T, Z)\right]
}
\right).
\]
By changing \( \eta \), we can assess the potential impact of deviations from the transportability assumption (\( \eta=0\)) on the parameter estimates. Given the estimated \(\hat{\gamma}_{n'}\) and the chosen value of the sensitivity parameter \( \eta \), estimation proceeds by maximizing the following pseudo-likelihood:
\begin{equation}
\ell_n(\theta;\hat{\gamma}_{n'}, \eta) =
\sum_{i=1}^{n} \left(
    \sum_{j=1}^{k} \Delta^*_{ij} \log 
    \left\{
        \sum_{h=1}^{k} \exp\left[\phi_h(X_i) + \beta_h^T Z_i\right] 
        \phi_h'(X_i) \, \pi_{jh}^*(X_i, Z_i; \hat{\gamma}_{n'},\eta)
    \right\}
    - \sum_{j=1}^{k} \exp\left[\phi_j(X_i) + \beta_j^T Z_i\right]
\right),
\label{eq:logPL_p12_p1}
\end{equation}
where $\pi_{jh}^*(X_i, Z_i; \hat{\gamma}_{n'},\eta)$ is the estimate of $P(C^* = j \mid C = h, S=1, T = t, Z = z)$ under the chosen $\eta$.

\section{Simulation study}\label{sec:sim_p1}

A series of simulation studies were carried out to evaluate the performance of the proposed estimator in finite samples. These simulations were based on a competing risks model with two causes of failure. One covariate $Z$ was used in the simulation, which was simulated from a normal distribution with mean 1 and standard deviation 1. Failure times $T_1$ and $T_2$ were generated from proportional hazards models:
\[
\lambda_1(t|Z) = 0.5 \exp(0.6Z),
\]
and
\[
\lambda_2(t|Z) = 0.5 \exp( 2t)\exp(0.3Z),
\]
respectively. The failure rates for the two types of failure were, on average, $46.9-47.2\%$ and $31.8-32.1\%$ failures from causes 1 and 2, respectively. The censoring times were generated independently from a uniform distribution on the interval $(0, 2)$. This censoring mechanism led to approximately $20.9-21.3\%$ right-censoring in the datasets. 

To resemble our motivating EA-IeDEA study, we assumed a unidirectional misclassification mechanism, where only cause 2 could be misclassified as cause 1, but not vise versa. We considered three scenarios for the logit model for the misclassification probability:
\begin{itemize}
    \item[Scenario 1:] \(\text{logit}\{\pi_{12}^*(W; \gamma_0)\} = \gamma_0 -0.7T + 0.8Z, \)
    \item[Scenario 2:] \(\text{logit}\{\pi_{12}^*(W; \gamma_0)\} = \gamma_0 -0.7\text{log}(T) + 0.8Z, \)
    \item[Scenario 3:] \(\text{logit}\{\pi_{12}^*(W; \gamma_0)\} = \gamma_0 -0.7T^2 + 0.8Z, \)
\end{itemize}
where $W = (T, Z)$ and $\gamma_0 \in \{-2.0, -1.8, -1.5\}$, leading to approximately $0.191$, $0.246$, and $0.284$ misclassification probabilities, respectively.

For our simulation experiments, we generated 1,000 datasets for each scenario, with main study sample sizes $n = 400, 600,$ and $800$. For simplicity, and following \cite{ha2015semiparametric}, the misclassification probabilities were considered to be fixed in the simulation studies. For analysis purposes, these misclassification probabilities were assumed to follow the logistic regression model: \(\text{logit}\{\pi_{12}^*(W; \gamma)\} = \gamma_0 +\gamma_1T + \gamma_2Z\), which was correct under scenario 1 and misspecified under scenario 2 and scenario 3. The latter two scenarions, provide numerical evidence for the robustness of the proposed approach under a misspecified model for the misclassification probabilities. In the simulation experiments we evaluated  the performance of the proposed method and demonstrated its efficiency advantage over the HT-WM approach \cite{ha2015semiparametric}. We focused on this particular method, as it is reported to be the most efficient among the estimating-equations-based methods, which do not impose parametric assumptions on the baseline cause-specific hazard functions. Given that this simulation study did not aim at evaluating the standard error estimator for the HT-WM method and in an effort to avoid additional computation time, we do not report average standard errors and coverage probabilities for this method.

The simulation results are summarized in Tables~\ref{table:simulation_results}, \ref{table:simulation_results_log_linear}, and \ref{table:simulation_results_quadratic_linear}, presenting results from Scenarios 1, 2, and 3, respectively. The proposed methodology exhibited virtually unbiased estimates, while the average of the standard error estimates were close to the corresponding Monte Carlo standard deviation (MCSD) of the regression coefficient estimates. In addition, the empirical coverage probabilities of the 95\% confidence intervals were close to the nominal level. These results provide numerical evidence for the validity of the proposed methodology.  

The proposed method consistently showed superior performance in terms of efficiency (smaller MCSD) compared to the HT-WM method, across all sample sizes and misclassification rates. This phenomenon was more profound at smaller sample sizes and higher missclasification rates. In addition, the proposed approach resulted in estimates with slightly smaller bias across all scenarios and sample sizes. The superiority of the performance of the proposed approach was retained even under a misspecified missclasification model (scenario 2 and scenario 3). For example, in scenario 3 with $p_m = 28.4\%$ and $n = 400$, the relative efficiency was as high as $1.637$, indicating that the HT-WM method exhibited over 60\% increased variance, compared to our method. 

\begin{table}[ht]
\centering
\footnotesize
\begin{tabular}{c c c c c c c c c c c c c}
\hline
$p_m (\%)$ & $n$ & Method & \multicolumn{5}{c}{$\beta_1$} & \multicolumn{5}{c}{$\beta_2$} \\
\cmidrule(lr){4-8}
\cmidrule(lr){9-13}
& & & Bias (\%) & MCSD & RE & ASE & CP & Bias (\%) & MCSD & RE & ASE & CP \\
\hline
19.1 & 400 & Proposed & 1.880 & 0.133 & 1 & 0.141 & 0.955 & 0.202 & 0.106 & 1 & 0.113 & 0.954 \\
     &     & HT-WM    & -1.525 & 0.145 & 1.186 &        &       & 1.783 & 0.106 & 1.001 &        &       \\
     & 600 & Proposed & -0.102 & 0.109 & 1 & 0.118 & 0.966 & 1.398 & 0.093 & 1 & 0.096 & 0.954 \\
     &     & HT-WM    & -1.448 & 0.109 & 1.002 &        &       & 2.524 & 0.094 & 1.021 &        &       \\
     & 800 & Proposed & 0.232 & 0.100 & 1 & 0.108 & 0.960 & 1.742 & 0.083 & 1 & 0.088 & 0.961 \\
     &     & HT-WM    & -1.188 & 0.100 & 1.001 &        &       & 2.795 & 0.083 & 1.005 &        &       \\
24.6 & 400 & Proposed & 2.091 & 0.141 & 1 & 0.150 & 0.956 & -0.097 & 0.110 & 1 & 0.118 & 0.954 \\
     &     & HT-WM    & -2.163 & 0.160 & 1.290 &        &       & 2.037 & 0.112 & 1.037 &        &       \\
     & 600 & Proposed & -0.893 & 0.115 & 1 & 0.124 & 0.952 & 2.673 & 0.095 & 1 & 0.099 & 0.959 \\
     &     & HT-WM    & -3.019 & 0.118 & 1.063 &        &       & 3.777 & 0.097 & 1.043 &        &       \\
     & 800 & Proposed & 0.440 & 0.103 & 1 & 0.114 & 0.965 & 1.226 & 0.086 & 1 & 0.092 & 0.967 \\
     &     & HT-WM    & -1.063 & 0.104 & 1.022 &        &       & 2.566 & 0.086 & 1.008 &        &       \\
28.4 & 400 & Proposed & 1.578 & 0.152 & 1 & 0.163 & 0.957 & 0.107 & 0.118 & 1 & 0.124 & 0.963 \\
     &     & HT-WM    & -4.064 & 0.185 & 1.468 &        &       & 3.594 & 0.121 & 1.050 &        &       \\
     & 600 & Proposed & -0.492 & 0.127 & 1 & 0.139 & 0.964 & 1.044 & 0.103 & 1 & 0.108 & 0.955 \\
     &     & HT-WM    & -3.637 & 0.140 & 1.212 &        &       & 2.859 & 0.107 & 1.074 &        &       \\
     & 800 & Proposed & 0.374 & 0.111 & 1 & 0.127 & 0.965 & 0.907 & 0.092 & 1 & 0.099 & 0.961 \\
     &     & HT-WM    & -1.933 & 0.121 & 1.183 &        &       & 2.096 & 0.094 & 1.043 &        &       \\
\hline
\end{tabular}
\raggedright
\footnotesize
$p_m (\%)$: percent of misclassification; MCSD: Monte Carlo standard deviation; RE: relative efficiency; ASE: average standard error; CP: coverage probability
\vspace{0.1cm}
\caption{Simulation results for $\beta_1$ and $\beta_2$ under Scenario 1 with correct misclassification model specification, comparing the proposed method to the HT-WM method across varying misclassification rates and sample sizes.}
\label{table:simulation_results}
\end{table}


\begin{table}[ht]
\centering
\footnotesize
\begin{tabular}{c c c c c c c c c c c c c}
\hline
$p_m (\%)$ & $n$ & Method & \multicolumn{5}{c}{$\beta_1$} & \multicolumn{5}{c}{$\beta_2$} \\
\cmidrule(lr){4-8}
\cmidrule(lr){9-13}
& & & Bias (\%) & MCSD & RE & ASE & CP & Bias (\%) & MCSD & RE & ASE & CP \\
\hline
19.1 & 400 & Proposed & 1.916 & 0.133 & 1 & 0.142 & 0.958 & 0.128 & 0.107 & 1 & 0.114 & 0.952 \\
     &     & HT-WM    & -1.148 & 0.145 & 1.176 &        &       & 2.116 & 0.107 & 1.002 &        &       \\
     & 600 & Proposed & -0.160 & 0.109 & 1 & 0.119 & 0.966 & 1.438 & 0.093 & 1 & 0.096 & 0.950 \\
     &     & HT-WM    & -1.683 & 0.109 & 1.002 &        &       & 2.279 & 0.094 & 1.021 &        &       \\
     & 800 & Proposed & 0.371 & 0.100 & 1 & 0.108 & 0.964 & 1.528 & 0.083 & 1 & 0.089 & 0.961 \\
     &     & HT-WM    & -0.956 & 0.102 & 1.031 &        &       & 2.230 & 0.084 & 1.026 &        &       \\
24.6 & 400 & Proposed & 2.128 & 0.142 & 1 & 0.151 & 0.957 & -0.147 & 0.112 & 1 & 0.119 & 0.954 \\
     &     & HT-WM    & -2.290 & 0.164 & 1.344 &        &       & 1.830 & 0.114 & 1.029 &        &       \\
     & 600 & Proposed & -0.816 & 0.116 & 1 & 0.126 & 0.961 & 2.442 & 0.096 & 1 & 0.101 & 0.957 \\
     &     & HT-WM    & -2.832 & 0.119 & 1.055 &        &       & 3.786 & 0.098 & 1.052 &        &       \\
     & 800 & Proposed & 0.306 & 0.104 & 1 & 0.115 & 0.966 & 1.398 & 0.087 & 1 & 0.093 & 0.972 \\
     &     & HT-WM    & -1.379 & 0.107 & 1.051 &        &       & 2.521 & 0.089 & 1.049 &        &       \\
28.4 & 400 & Proposed & 1.572 & 0.156 & 1 & 0.166 & 0.954 & -0.226 & 0.119 & 1 & 0.126 & 0.961 \\
     &     & HT-WM    & -4.120 & 0.186 & 1.429 &        &       & 2.673 & 0.122 & 1.049 &        &       \\
     & 600 & Proposed & -0.632 & 0.129 & 1 & 0.139 & 0.961 & 1.229 & 0.103 & 1 & 0.109 & 0.955 \\
     &     & HT-WM    & -3.735 & 0.142 & 1.221 &        &       & 2.884 & 0.108 & 1.098 &        &       \\
     & 800 & Proposed & 0.056 & 0.113 & 1 & 0.128 & 0.966 & 1.295 & 0.094 & 1 & 0.100 & 0.958 \\
     &     & HT-WM    & -1.892 & 0.121 & 1.153 &        &       & 2.338 & 0.096 & 1.047 &        &       \\
\hline
\end{tabular}
\raggedright
\footnotesize
$p_m (\%)$: percent of misclassification; MCSD: Monte Carlo standard deviation; RE: relative efficiency; ASE: average standard error; CP: coverage probability
\vspace{0.1cm}
\caption{Simulation results for $\beta_1$ and $\beta_2$ under Scenario 2 with misclassification model misspecification, comparing the proposed method to the HT-WM method across varying misclassification rates and sample sizes.}
\label{table:simulation_results_log_linear}
\end{table}

\begin{table}[ht]
\centering
\footnotesize
\begin{tabular}{c c c c c c c c c c c c c}
\hline
$p_m (\%)$ & $n$ & Method & \multicolumn{5}{c}{$\beta_1$} &  \multicolumn{5}{c}{$\beta_2$} \\
\cmidrule(lr){4-8}
\cmidrule(lr){9-13}
& & & Bias (\%) & MCSD & RE & ASE & CP & Bias (\%) & MCSD & RE & ASE & CP \\
\hline
19.1 & 400 & Proposed & 1.887 & 0.137 & 1 & 0.143 & 0.957 & 0.266 & 0.109 & 1 & 0.115 & 0.953 \\
     &     & HT-WM    & -1.572 & 0.151 & 1.209 &        &       & 2.064 & 0.110 & 1.014 &        &       \\
     & 600 & Proposed & -0.655 & 0.112 & 1 & 0.121 & 0.962 & 2.070 & 0.094 & 1 & 0.097 & 0.958 \\
     &     & HT-WM    & -2.522 & 0.115 & 1.045 &        &       & 3.046 & 0.094 & 1.010 &        &       \\
     & 800 & Proposed & 0.398 & 0.102 & 1 & 0.111 & 0.969 & 1.216 & 0.085 & 1 & 0.090 & 0.961 \\
     &     & HT-WM    & -1.208 & 0.103 & 1.021 &        &       & 2.098 & 0.086 & 1.023 &        &       \\
24.6 & 400 & Proposed & 2.026 & 0.141 & 1 & 0.153 & 0.958 & -0.276 & 0.114 & 1 & 0.120 & 0.955 \\
     &     & HT-WM    & -2.954 & 0.172 & 1.479 &        &       & 2.172 & 0.116 & 1.026 &        &       \\
     & 600 & Proposed & -0.809 & 0.120 & 1 & 0.128 & 0.950 & 2.182 & 0.098 & 1 & 0.102 & 0.953 \\
     &     & HT-WM    & -3.293 & 0.127 & 1.105 &        &       & 3.693 & 0.101 & 1.055 &        &       \\
     & 800 & Proposed & 0.302 & 0.104 & 1 & 0.117 & 0.969 & 1.432 & 0.088 & 1 & 0.095 & 0.963 \\
     &     & HT-WM    & -1.554 & 0.110 & 1.128 &        &       & 2.359 & 0.090 & 1.043 &        &       \\
28.4 & 400 & Proposed & 1.211 & 0.159 & 1 & 0.169 & 0.962 & 0.073 & 0.121 & 1 & 0.128 & 0.966 \\
     &     & HT-WM    & -6.081 & 0.204 & 1.637 &        &       & 4.520 & 0.125 & 1.076 &        &       \\
     & 600 & Proposed & -0.975 & 0.130 & 1 & 0.143 & 0.953 & 1.933 & 0.105 & 1 & 0.110 & 0.958 \\
     &     & HT-WM    & -4.710 & 0.147 & 1.279 &        &       & 4.374 & 0.110 & 1.094 &        &       \\
     & 800 & Proposed & -0.024 & 0.118 & 1 & 0.130 & 0.970 & 1.466 & 0.095 & 1 & 0.101 & 0.962 \\
     &     & HT-WM    & -2.681 & 0.132 & 1.256 &        &       & 2.909 & 0.098 & 1.064 &        &       \\
\hline
\end{tabular}
\raggedright
\footnotesize
$p_m (\%)$: percent of misclassification; MCSD: Monte Carlo standard deviation; RE: relative efficiency; ASE: average standard error; CP: coverage probability
\vspace{0.1cm}
\caption{Simulation results for $\beta_1$ and $\beta_2$ under Scenario 3 with misclassification model misspecification, comparing the proposed method to the HT-WM method across varying misclassification rates and sample sizes.}
\label{table:simulation_results_quadratic_linear}
\end{table}

\section{HIV data analysis}\label{sec:dataapp_p1}

We applied the proposed methodology to electronic health record data from the EA-IeDEA regional consortium. In particular, we were interested in modeling the cause-specific hazard of gap in HIV care and death in order to identify risk factors for these two causes of failure. The covariates of interest were sex, age at antiretroviral therapy (ART) initiation, and CD4 (cluster of differentiation 4) count at ART initiation. Data were obtained from two EA-IeDEA programs, the Family AIDS Care \& Education Services (FACES) and the Academic Model Providing Access to Healthcare (AMPATH). The FACES cohort was the main study population, while the AMPATH cohort was used solely to provide externally estimated misclassification probabilities. AMPATH investigators employ a double sampling design where a small sample of individuals who were identified as lost to clinic is intensively outreached in the community and their true vital status is actively ascertained by field workers \citep{yiannoutsos2008sampling, an2009need}. These data were leveraged to estimate the probability of an  unreported death as a function of covariates. This estimate was then used in the proposed methodology to adjust for misclassification due to death under-reporting. Data from both programs were collected between 2001 and 2011. Finally, the data from FACES included 3,886 HIV-infected adult patients, while the AMPATH dataset involved 63,890 patients.

\begin{table}[ht]
\centering
\captionsetup{justification=centering} 
\scriptsize
\begin{tabular}{lcccc}
\hline
& & \multicolumn{2}{c}{\textbf{East Africa IeDEA Program}} \\ \cline{3-4} \textit{Variable} & Total, N=67776 & AMPATH, N=63890 & FACES, N=3886 & p-value \\
\hline
\textbf{Sex, n(\%)} & & & & \\
\quad Female           & 44506 (65.7) & 41944 (65.7) & 2562 (66.0) & 0.736 \\
\quad Male             & 23270 (34.3) & 21946 (34.3) & 1324 (34.0) &       \\
\textbf{Age at ART initiation, in years} & & & & \\
\quad Mean (SD)         & 38 (9.9)     & 38.2 (9.8)   & 34 (9.7)    & $<$.001 \\
\quad Median (IQR)      & 36.7 (30.8 -- 44.1) & 37 (31.1 -- 44.3) & 32.2 (27 -- 39.2) & \\
\textbf{Time since ART initiation, in months} & & & & \\
\quad Mean (SD)         & 21.8 (20.1)  & 22.3 (20.4)  & 14.7 (11.1) & $<$.001 \\
\quad Median (IQR)      & 15.3 (5.5 -- 32.2) & 15.7 (5.5 -- 33.3) & 11.5 (5.6 -- 22.3) & \\
\textbf{CD4 count at ART initiation, in $\text{cells}/\mu l$} & & & & \\
\quad Mean (SD)         & 212.5 (182)  & 213.1 (183.1) & 202.8 (163.7) & $<$.001 \\
\quad Median (IQR)      & 182 (81 -- 287.9) & 182 (81 -- 288) & 182 (83 -- 282.8) & \\
\textbf{Observed Cause of Failure, n(\%)} & & & & \\
\quad Censored          & 34983 (51.6) & 32711 (51.2) & 2272 (58.5) & $<$.001 \\
\quad Observed Death     & 2792 (4.1)   & 2719 (4.3)   & 73 (1.9)    &       \\
\quad Observed Loss to Clinic & 30001 (44.3) & 28460 (44.5) & 1541 (39.6) & \\
\textbf{Cause of Failure after Outreach, n(\%)} & & & & \\
\quad Death after outreach &  & 1143 (27.0) &  & \\
\quad Loss to clinic after outreach &  & 3095 (73.0) &  & \\
\textbf{Double Sampling at AMPATH, n(\%)} & & & & \\
\quad Total in double sampling &  & 4238 (100.0) &  & \\
\hline
\end{tabular}
\parbox[b]{0.73\linewidth}{\raggedright\scriptsize\textit{P-values were obtained from two-sample t-tests for continuous variables and chi-squared tests for categorical variables, comparing AMPATH and FACES. Statistical significance was assessed at the $\alpha = 0.05$ level.}}
\caption{Descriptive statistics for the EA-IeDEA study sample overall and by program.}
\label{tab:patient_characteristics}
\end{table}

Table~\ref{tab:patient_characteristics} presents the descriptive characteristics for the EA-IeDEA study sample overall and by program. Using data from AMPATH, among the 28,460 patients classified as having a gap in care, double sampling (i.e., patient outreach) was carried out in a subset of 4,238 (14.9\%), revealing that 1,143 (27\%) had actually died. This reclassification increased the number of recorded deaths from 2,719 to 3,862, indicating that 29.6\% of all deaths were originally misclassified as gaps in care. This indicates a substantial death under-reporting issue. To account for this cause of failure misclassification, we first estimated the probability that a true death was misclassified as gap in care (i.e., an unreported death) using a logistic regression model \citep{mpofu2020pseudo}. The model included sex (male versus female), age at ART initiation (in years, centered), CD4 count at ART initiation (in square-root form), and time contributed to the study (in months, modeled piecewise linearly) as covariates:
\begin{equation}
\label{eq:miscl_model}
\begin{aligned}
\text{logit}\{\pi^*_{12}(W, \gamma_0)\}
={} & \gamma_0 
+ \gamma_1 \, I(\text{Sex} = \text{"male"})
+ \gamma_2 \, \text{Age}_{\text{center}}
+ \gamma_3 \, \sqrt{\text{CD4}}
+ \gamma_4 \, T \\
& + \gamma_5 \, I(3 \le T < 6)\,(T - 3)
+ \gamma_6 \, I(6 \le T < 12)\,(T - 6)
+ \gamma_7 \, I(T \ge 12)\,(T - 12),
\end{aligned}
\end{equation}
where $T$ is the time from ART initiation to any cause of failure. The piecewise constant model for time was used to relax the linearity assumption of time on the logit scale \cite{mpofu2020pseudo}. The parameter estimates $\hat{\gamma}_{n'}$, together with their standard errors and p-values, are reported in Table~\ref{tab:miscl_model}. The estimated misclassification probabilities were then incorporated into the proposed pseudolikelihood method to adjust for misclassification when estimating cause-specific hazards.

\begin{table}[htbp]
\centering
\scriptsize
\begin{tabular}{lcccc}
\hline
\textbf{Covariate} & $\hat{\gamma}_{n'}$ & \textbf{SE} & \textbf{p-value} & \textbf{OR (95\% CI)} \\
\hline
Intercept & 0.0119 & 0.0773 & 0.878 & --- \\
Male (vs Female) & $-0.2058$ & 0.0633 & 0.001 & 0.814 (0.719, 0.922) \\
Centered age (years) & 0.0041 & 0.0030 & 0.176 & 1.004 (0.998, 1.010) \\
$\sqrt{\text{CD4  }}(\text{cells}/\mu l)$ & 0.0077 & 0.0058 & 0.183 & 1.008 (0.996, 1.019) \\
Time since ART initiation (months) & & & & \\
\quad 0--3 months & 0.1768 & 0.0103 & $<0.001$ & 1.193 (1.170, 1.218) \\
\quad 3--6 months             & 0.3133 & 0.0442 & $<0.001$ & 1.368 (1.254, 1.492) \\
\quad 6--12 months            & $-0.1005$ & 0.0261 & $<0.001$ & 0.904 (0.859, 0.952) \\
\quad $\ge$12 months          & $-0.2048$ & 0.0127 & $<0.001$ & 0.815 (0.795, 0.835) \\
\hline
\end{tabular}
\vspace{0.1cm}
\parbox[b]{0.60\linewidth}{\raggedright\scriptsize\textit{OR: odds ratio; 95\% CI: 95\% confidence interval}}

\caption{Estimated missclasification model: logistic regression for the probability of death among those originally classified as having a gap in care (unreported death) at AMPATH.}
\label{tab:miscl_model}
\end{table}

\begin{table}[ht]
\centering
\small
\begin{tabular}{lcc}
\hline
 & \multicolumn{2}{c}{\textit{\textbf{Death while in care}}} \\ \cline{2-3}
\textbf{Covariate} & \textbf{Proposed} & \textbf{Naïve} \\
 & \textbf{HR (95\% CI)} & \textbf{HR (95\% CI)} \\
\hline
Age (10 years ↑) 
& 1.063 (0.677, 1.684) & 1.036 (0.804, 1.337) \\

Male (vs Female)
& 1.403 (0.735, 2.670) & 1.696 (1.073, 2.680)$^*$ \\

CD4 (100 cells/\(\mu\)l ↑)
& 0.669 (0.437, 1.024) & 0.725 (0.543, 0.966)$^*$ \\
\hline
\\[-0.3cm]
 & \multicolumn{2}{c}{\textit{\textbf{Gap in HIV care}}} \\ \cline{2-3}
\textbf{Covariate} & \textbf{Proposed} & \textbf{Naïve} \\
 & \textbf{HR (95\% CI)} & \textbf{HR (95\% CI)} \\
\hline
Age (10 years ↑) 
& 0.771 (0.700, 0.849)$^{***}$ & 0.791 (0.744, 0.841)$^{**}$ \\

Male (vs Female)
& 1.095 (0.931, 1.288) & 1.085 (0.973, 1.210) \\

CD4 (100 cells/\(\mu\)l ↑)
& 0.945 (0.883, 1.013) & 0.935 (0.897, 0.974)$^{**}$ \\
\hline
\end{tabular}

\parbox[b]{0.60\linewidth}{\raggedright\scriptsize\textit{HR: hazard ratio; CI: confidence interval; $^*$\(p < 0.05\); $^{**}$\(p < 0.01\); $^{***}$\(p < 0.001\).}}
\vspace{0.15cm}
\caption{Second step of the data analysis for the EA-IeDEA study: hazard ratios and 95\% confidence intervals comparing the proposed method ($\eta=0$) and the naïve method for death and gap in HIV care.}
\label{tab:proposedvsnaive}
\end{table}

In the second step of the analysis, which constituted the main analysis, we fitted cause-specific proportional hazards models of the form:
\[
\lambda_j(t; Z) = \lambda_{0,j}(t)
\exp\{\beta_{1,j} I(\text{Sex} = \text{``male''})
+ \beta_{2,j} \text{Age}
+ \beta_{3,j} \text{CD4}\},
\]
for each event type ($j = 1, 2$). The models were estimated using the proposed B-spline-based sieve pseudolikelihood approach together with the parameter estimates $\hat{\gamma}_{n'}$ from the first step. Variance estimation was performed according to Algorithm~\ref{alg:var} using $B=100$ bootstrap replications. 

Table~\ref{tab:proposedvsnaive} presents the hazard ratios and 95\% confidence intervals for the cause-specific hazards of death and gap in care, comparing the proposed method with the na\"{i}ve approach (HT-WM). For death, age at ART initiation remained a positive but non-significant predictor under both methods, with similar estimates between the proposed (HR = 1.063, 95\% CI: 0.688, 1.643) and naïve (HR = 1.036, 95\% CI: 0.804, 1.337) analyses. In contrast, the effect of sex became weaker after adjusting for misclassification, with the hazard ratio decreasing from 1.696 (95\% CI: 1.073, 2.680) in the naïve model to 1.403 (95\% CI: 0.717, 2.745) under the proposed method, losing statistical significance. Similarly, CD4 at ART initiation was negatively associated with the hazard of death under both approaches, but the association was no longer statistically significant after adjustment, with the proposed method producing a slightly stronger protective effect (HR = 0.669, 95\% CI: 0.414, 1.080) compared with the naïve model (HR = 0.725, 95\% CI: 0.543, 0.966). For gap in care, older age at ART initiation remained strongly associated with a lower hazard under both methods, with very similar estimates between the proposed (HR = 0.771, 95\% CI: 0.702, 0.847) and naïve (HR = 0.791, 95\% CI: 0.744, 0.841) analyses. This effect remained statistically significant in both cases. The effect of sex was consistently positive but not statistically significant under either method. Finally, CD4 at ART initiation showed a negative association with gap in care under both approaches. However, the association was no longer statistically significant under the proposed method (HR = 0.945, 95\% CI: 0.873, 1.023), whereas the naïve analysis indicated a modest, yet statistically significant, negative association (HR = 0.935, 95\% CI: 0.897, 0.974).

\begin{figure}[ht]
\centering
\includegraphics[width=0.9\textwidth]{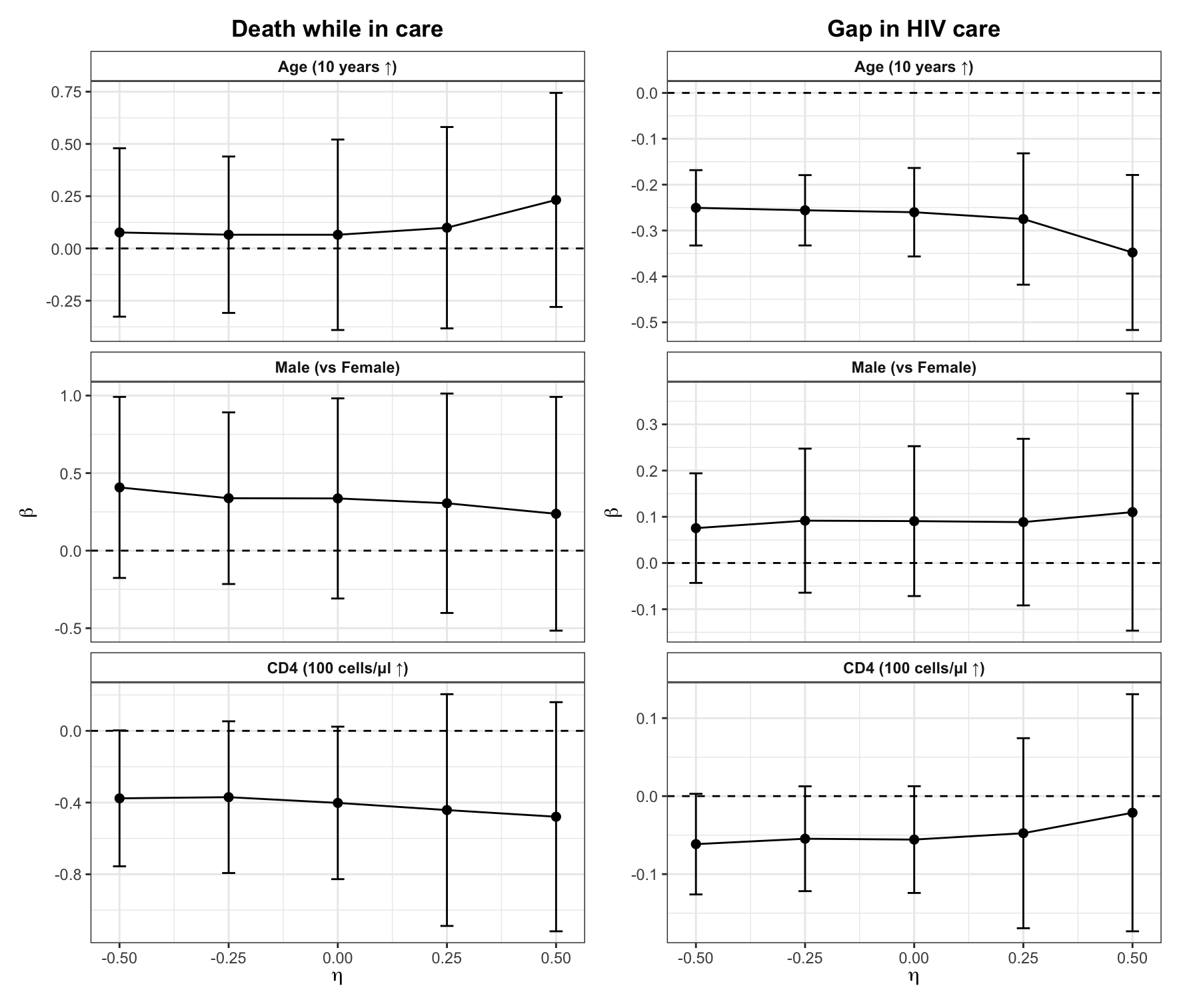}
\caption{Sensitivity analysis results for the proposed method across values of the sensitivity parameter $\eta$. The plots display the estimated regression coefficients ($\beta$) and corresponding 95\% confidence intervals for death while in care and gap in HIV care. The dashed horizontal line represents $\beta = 0$.}
\label{fig:sensitivity}
\end{figure}

To assess the effects of potential violations of the transportability assumption, we applied the sensitivity analysis approach presented in Section~\ref{sec:sens_p1}. Specifically, we varied the sensitivity parameter $\eta$ over the set $\{-0.5, -0.25, 0, 0.25, 0.5\}$. This means that the assumed baseline odds ratio of death under-reporting for the main study (FACES) vs. the external validation study (AMPATH) ranged from $\exp(-0.5)\approx 0.61$ to $\exp(0.5)\approx 1.65$. Figure~\ref{fig:sensitivity} shows how the regression coefficients ($\beta$) change across different values of the sensitivity parameter. Overall, the estimates remain relatively stable across $\eta$, suggesting that the analysis results are robust against moderate violations of this assumption.

\begin{figure}[htbp]
    \centering
    \includegraphics[width=0.7\textwidth]{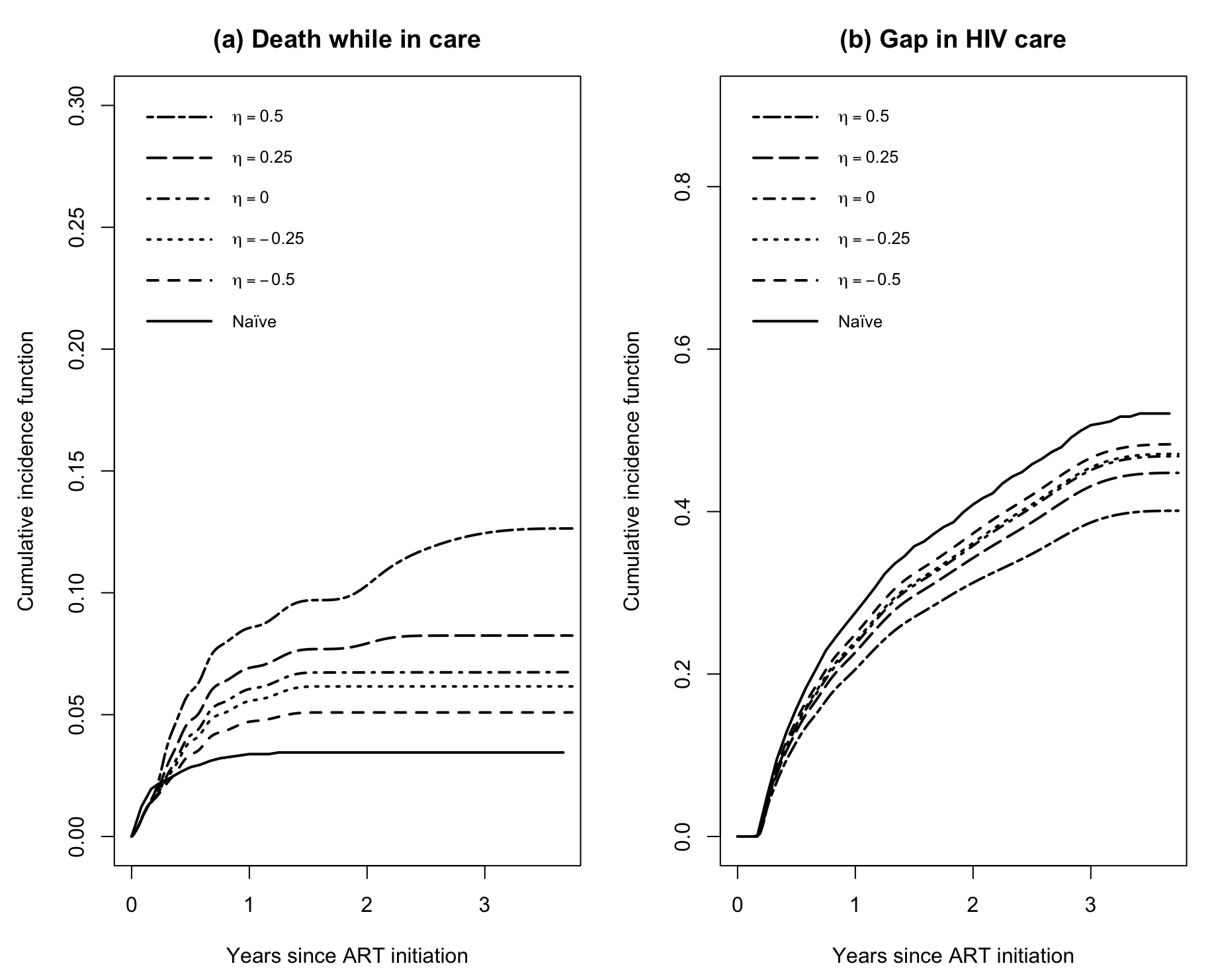}
    \caption{\small Predicted cumulative incidence functions of (a) death and (b) gap in HIV care for a 45-year-old male with a CD4 count of 150~cells/$\mu$L at ART initiation, comparing the naïve model (solid line) with the proposed method under different $\eta$ values (dashed lines).}
    \label{fig:cifs}
\end{figure}

To illustrate the proposed approach for risk prediction, we presented the predicted cumulative incidence functions for (a) death while in care and (b) gap in HIV care for a 45-year-old male with a CD4 count of 150~cells/$\mu$L at ART initiation, as shown in Figure~\ref{fig:cifs}. The naïve analysis underestimates the cumulative incidence of death and overestimates the probability of gap in care compared to the proposed method. These results are as expected given that misclassification here is a consequence of death under-reporting \citep{brinkhof2010adjusting}. Additionally, these results suggest that a 45-year-old male patient with a CD4 count of 150 cells/$\mu$l at ART initiation has a relatively high risk of disengaging from care over time. This indicates the need for improved patient support and follow-up to encourage continued care. In addition, the risk of death is also substantial, suggesting that closer monitoring and more intensive care may be needed. Figure~\ref{fig:cifs} also depicts the predicted cumulative incidence functions based on the sensitiviy analysis (i.e., $\eta\neq 0$). Based on this analysis, it can be seen that, unlike the estimated regression parameter estimates $\hat{\beta}_n$, the predicted cumulative incidence functions appear to be somewhat sensitive to violations of the transportability assumption.

\section{Discussion}\label{sec:disc_p1}

In this paper, we proposed a semiparametric regression methodology for competing risks data subject to cause of failure misclassification.  This approach leverages externally estimated misclassification probabilities, which are allowed to depend on covariates and time \citep{mpofu2020pseudo}. These estimates are being plugged into a B-spline-based likelihood function which adjusts for cause of failure misclassification. This function is maximized to obtain the parameter estimates of interest. By utilizing empirical process theory, we established the consistency of the proposed estimator. Variance estimation is performed using a bootstrap-based algorithm (Algorithm~\ref{alg:var}) that accounts for both the variability in the main study sample and the variability of the externally estimated miclassification parameter $\hat{\gamma}_{n'}$. Importantly, this algorithm does not require access to the data from the external validation study, depending only on the estimate $\hat{\gamma}_{n'}$ and its estimated variance matrix $\hat{\Omega}_{n'}$. The proposed method relies on the transportability assumption, that is  the misclassification probabilities in the main study are equal to those in the external validation study. To address potential violations of this assumption, we proposed a sensitivity analysis approach. Simulation studies demonstrated that the method works well with finite samples, and that it achieves substantially higher statistical efficiency compared to existing approaches \cite{ha2015semiparametric}, particularly under high misclassification rates, small sample sizes, and misspecified misclassification models. Coverage probabilities of the confidence intervals were at the nominal level across all scenarios considered, a result that further supports the validity of the proposed method and resulting inference. The method was applied to HIV care data from the EA-IeDEA regional consortium. Based on this analysis, our approach yielded different estimates compared to the na\"ive analysis that ignores misclassification of the cause of failure. Importantly, our methodology is straightforward to implement in practice using the R code available on GitHub (\url{https://github.com/tbalanos/misclassified-competing-risks-bssmle}).

To the best of our knowledge, the issue of semiparametric regression analysis of misclassified competing risks data using misclassification probabilities has primarily been addressed by Ha and Tsodikov \cite{ha2015semiparametric}. Our method extends beyond this approach by providing a robust, likelihood-based, semiparametric approach that does not impose parametric assumptions on the baseline cause-specific hazard functions. By virtue of our method being likelihood based, it resulted in substantial statistical efficiency gains over the semiparametric estimating equations HT-WM approach \cite{ha2015semiparametric}. Furthermore, unlike the HT-WM approach, we explicitly account for the additional uncertainty in the externally estimated misclassification probabilities. In addition, the HT-WM approach was found to be computationally intensive in our simulation studies. For example, under scenario 1 with \(n = 800\), the proposed method completed point estimation in approximately 1.84 seconds on average, compared with 38.99 seconds for HT-WM, an approximately 22-fold reduction in computation time. This improvement is particularly appealing feature when analyzing modern large datasets. Finally, to the best of our knowledge, there is no software available in common packages to readily implement the HT-WM approach. To the best of our knowledge, our R function (\url{https://github.com/tbalanos/misclassified-competing-risks-bssmle}) is the first to readily fit semiparametric proportional cause-specific hazards models with misclassified competing risks data.

In conclusion, this work presents a robust and computationally efficient framework for analyzing competing risks data subject to cause of failure misclassification. By leveraging external validation data, the approach yields more accurate regression parameter estimates compared to previous methods \cite{ha2015semiparametric}. In addition, variance estimation in our approach does not require access to the external validation data, avoiding confidentiality issues or other issues related to the access of these data. 
Future work may extend the methodology to the more general class of semiparametric transformation models, which includes the proportional cause-specific hazards model as a special case, or to more complex multistate models under misclassification of the states of the multistate process of interest. 

\par\null

\section*{Software}

An R illustrative example implementing the proposed method is publicly available at the GitHub repository: \url{https://github.com/tbalanos/misclassified-competing-risks-bssmle}.

\section*{Appendix}

\subsection*{Proof of Theorem~\ref{thm:consistency_p1}}

Empirical process theory techniques were used to establish the consistency of the proposed B-spline sieve maximum likelihood estimator. Here we use the standard empirical process theory notations 
\(
Pf = \int f dP = E\{f(D)\}\) and \(\mathbb{P}_n f = n^{-1} \sum_{i=1}^n f(D_i),
\)
where $D=(\Delta_1^*,\ldots, \Delta_k^*, X, Z)\in\mathcal{D}$ is a data point and $\mathcal{D}$ is the sample space. Next, define the class of functions $\mathcal{F}_1(\gamma_0)=\{\,l(\theta;\gamma_0):\theta\in\Theta_n\,\}$, where
\[
l(\theta;\gamma_0)=\sum_{j=1}^{k} \Delta^*_{ij} \log 
    \left\{
        \sum_{h=1}^{k} \exp\left[\phi_h(X_i) + \beta_h^T Z_i\right] 
        \phi_h'(X_i) \, \pi_{jh}^*(X_i, Z_i; \gamma_0)
    \right\}-  
    \sum_{j=1}^{k} \exp\left[\phi_j(X_i) + \beta_j^T Z_i\right],
\]
is the class of
log-likelihood functions indexed by the sieve parameter space
\[
\Theta_n=\times_{j=1}^{k}\big(\mathcal{B}_j \times \Phi_{n,j}\big),
\]
where the finite-dimensional parameters are $\beta_j\in\mathcal{B}_j$ and the infinite-dimensional parameters are
$\phi_j\in\Phi_{n,j}$, $j=1,\ldots,k$. The sieve
$\Phi_{n,j}$ consists of monotone B\mbox{-}spline basis functions on $[0,\tau]$.
Let $\Theta$ denote the true (non-sieve) parameter space. Consistency is proved under the $L_2$-metric $d(\cdot,\cdot)$ defined by
\[
d\!\big(\theta^{(1)},\theta^{(2)}\big)
=\Bigg(\sum_{j=1}^{k}\big\|\beta^{(1)}_{j}-\beta^{(2)}_{j}\big\|^{2}
\;+\;\sum_{j=1}^{k}\big\|\phi^{(1)}_{j}-\phi^{(2)}_{j}\big\|_{\Phi}^{2}\Bigg)^{1/2},
\]
for $\theta^{(1)},\theta^{(2)}\in\Theta$, where $\|\cdot\|$ denotes the Euclidean norm and the norm for the infinite-dimensional parameters is defined as
\[
\big\|\phi^{(1)}_{j}-\phi^{(2)}_{j}\big\|_{\Phi}^{2}
=E\!\left[\big\{\phi^{(1)}_{j}(T)-\phi^{(2)}_{j}(T)\big\}^{2}\right],\qquad j=1,\ldots,k. 
\]
Now, let $\theta_0$ denote the true parameter value. In this work, we assume the following conditions hold:
\begin{itemize}
    \item[C1.] \( E(ZZ^{'}) \) is non-singular, and \( Z \) is bounded, i.e., there exists a constant \( z_0 > 0 \) such that \( P( \| Z\| \leq z_0) = 1 \).
    \item[C2.] \( \beta_j \in \mathcal{B}_j \), where \( \mathcal{B}_j \) is a compact subset of \( \mathbb{R}^d \) for every \( j = 1, \ldots, k \).
    \item[C3.] \( \phi_{0,j} \in \Phi \), where \( \Phi \) is a function space including functions with bounded \( p \)th derivative on \( [0, \tau] \) for \( p \geq 1 \) and the first derivative of \( \phi_{0,j} \) is strictly positive and continuous on \( [0, \tau] \). This holds for all \( j = 1, \ldots, k \).
    \item[C4.] For all causes of failure \( j =1,\ldots,k\), there exists a constant \( c_0 > 0.5 \) such that, \(\pi^*_{jj}(T, Z; \gamma) \geq c_0\), almost surely.
\end{itemize}

Let $\mathbb{M}(\theta; \gamma_0) = P l(\theta; \gamma_0)$ and $\mathbb{M}_n(\theta; \hat{\gamma}_{n'}) = \mathbb{P}_n l(\theta; \hat{\gamma}_{n'})$, which leads to
\[
\mathbb{M}_n(\theta; \hat{\gamma}_{n'}) - \mathbb{M}(\theta; \gamma_0) = \mathbb{P}_n[l(\theta; \hat{\gamma}_{n'}) - l(\theta; \gamma_0)] + \big(\mathbb{M}_n(\theta; \gamma_0) - \mathbb{M}(\theta; \gamma_0)\big).
\]
Proving that \( d(\hat{\theta}_n, \theta_0) \xrightarrow{p} 0 \), requires verifying the following conditions:

\begin{enumerate}
    \item
    \(
    \sup_{\theta \in \Theta_n} \big| \mathbb{M}_n(\theta; \hat{\gamma}_{n'}) - \mathbb{M}(\theta; \gamma_0) \big| \xrightarrow{p} 0.
    \)

    \item
    \(
    \sup_{\theta : d(\theta, \theta_0) \geq \epsilon} \mathbb{M}(\theta; \gamma_0) < \mathbb{M}(\theta_0; \gamma_0).
    \)

    \item The sequence of estimators $\hat{\theta}_n$ satisfies
    \(
    \mathbb{M}_n(\hat{\theta}_n; \hat{\gamma}_{n'}) \geq \mathbb{M}_n(\theta_0; \hat{\gamma}_{n'}) - o_p(1).
    \)
\end{enumerate}
For condition 1, we have
\begin{equation}
\sup_{\theta \in \Theta_n} \big|\mathbb{M}_n(\theta; \hat{\gamma}_{n'}) - \mathbb{M}(\theta; \gamma_0)\big| 
\leq \sup_{\theta \in \Theta_n} \big|\mathbb{P}_n[l(\theta; \hat{\gamma}_{n'}) - l(\theta; \gamma_0)]\big| 
+ \sup_{\theta \in \Theta_n} \big|\mathbb{M}_n(\theta; \gamma_0) - \mathbb{M}(\theta; \gamma_0)\big|, \label{condition1}
\end{equation}
Applying a second order Taylor expansion around $\gamma_0$ in the first term in the right side of \eqref{condition1}, we have that 
\[
l(\theta; \hat{\gamma}_{n'}) = l(\theta; \gamma_0) + \nabla_\gamma l(\theta; \gamma_0)^T (\hat{\gamma}_{n'} - \gamma_0) + \frac{1}{2} (\hat{\gamma}_{n'} - \gamma_0)^T \nabla_\gamma^2 l(\theta; \tilde{\gamma}_{n'}) (\hat{\gamma}_{n'} - \gamma_0),
\]
where \( \|\tilde{\gamma}_{n'} -\gamma_0\|\le\|\hat{\gamma}_{n'}-\gamma_0\| \) almost surely. By regularity conditions C1--C3, which imply the uniform boundnedness (in probability) of the derivatives $\nabla_\gamma l(\theta; \gamma_0)$ and $\nabla_\gamma^2 l(\theta; \tilde{\gamma}_{n'})$, for all $\tilde{\gamma}_{n'}$ in a neighborhood of $\gamma_0$, it follows that
\begin{eqnarray*}
\sup_{\theta \in \Theta_n} \big|\mathbb{P}_n[l(\theta; \hat{\gamma}_{n'}) - l(\theta; \gamma_0)]\big| &\leq& \|\hat{\gamma}_{n'}-\gamma_0\|O_p(1) + \|\hat{\gamma}_{n'}-\gamma_0\|^2O_p(1) \\
&=&o_p(1),
\end{eqnarray*}
as a consequence of the consistency of $\hat{\gamma}_{n'}$. For the second term in the right side of \eqref{condition1}, we focus on the class of functions $\mathcal{F}_1(\gamma_0)$. By the bracketing entropy for B-splines \cite{shen1994convergence}, Theorem 2.4.3 in \cite{vanderVaart1996weak}, and the fact that the loglikelihood function is continuous in $\theta$ along with Corollary 9.26 in \cite{kosorok2008introduction}, it follows that the class of functions $\mathcal{F}_1(\gamma_0)$ is Glivenko--Cantelli. This implies that 
\[
\sup_{\theta \in \Theta_n} \big|\mathbb{M}_n(\theta; \gamma_0) - \mathbb{M}(\theta; \gamma_0)\big|=o_p(1),
\]
and, thus, condition 1 is satisfied by \eqref{condition1}.

For condition 2, let $P_{\theta,\gamma_0}$ denote the probability measure for the observed data under the parameter value $\theta\in\Theta$. Then, the Gibb's inequality for the Kullback--Leibler divergence implies that
\[
\mathbb{M}(\theta;\gamma_0)\leq \mathbb{M}(\theta_0;\gamma_0),
\]
with equality if and only if $P_{\theta,\gamma_0}=P_{\theta_0,\gamma_0}$. Therefore, if $\mathbb{M}(\theta;\gamma_0)=\mathbb{M}(\theta_0;\gamma_0)$, then $P_{\theta,\gamma_0}=P_{\theta_0,\gamma_0}$, which implies that 
\begin{equation}
\exp\left[\phi_h(x) + \beta_h^T z\right] \phi_h'(x) \, \pi_{jh}^*(x, z; \gamma_0) = \exp\left[\phi_{0,h}(x) + \beta_{0,h}^T z\right] \phi_{0,h}'(x) \, \pi_{jh}^*(x, z; \gamma_0), \qquad h,j=1,\ldots,k, \label{misc_haz_eq}
\end{equation}
for all $(x,z)$ in the corresponding sample subspace. Next, let $\Pi^*(x,z)=[\pi_{jh}^*(x, z; \gamma_0)]_{1\leq j,h
\leq k}$ be the $k\times k$ matrix of the true classification probabilities $\pi_{jh}^*(x, z; \gamma_0)$, and 
\[
H(x,z;\theta)=\left(\exp\left[\phi_1(x) + \beta_1^T z\right] \phi_1'(x),\ldots,\exp\left[\phi_k(x) + \beta_k^T z\right] \phi_k'(x)\right)^T,
\]
for all $(x,z)$ in the sample subspace. Then \eqref{misc_haz_eq}, can be compactly expressed as
\begin{equation}
\Pi^*(x,z)\,H(x,z;\theta)=\Pi^*(x,z)\,H(x,z;\theta_0), \label{matrix_eq_1}
\end{equation}
In light of condition C4, $\pi_{jj}^*(x, z; \gamma_0)>\pi_{jh}^*(x, z; \gamma_0)$ for all $h\neq j$, and thus $\Pi^*(x,z)$ is strictly diagonally dominant. This implies that $\Pi^*(x,z)$ is a non-singular matrix and thus invertible. Denoting the inverse of this matrix by $\Pi^*(x,z)^{-1}$, and multiplying (on the left) both sides of \eqref{matrix_eq_1} by $\Pi^*(x,z)^{-1}$, leads to the conclusion that $H(x,z;\theta)=H(x,z;\theta_0)$, and thus,
\[
\exp\left[\phi_h(x) + \beta_h^T z\right] \phi_h'(x) = \exp\left[\phi_{0,h}(x) + \beta_{0,h}^T z\right] \phi_{0,h}'(x), \qquad h,j=1,\ldots,k,
\]
By conditions C1 and C3, this implies that $\beta_{j}=\beta_{0,j}$ and $\phi_{j}=\phi_{0,j}$, $j=1,\ldots,k$, and thus $\theta=\theta_0$. Therefore, we have shown that if $\mathbb{M}(\theta;\gamma_0)=\mathbb{M}(\theta_0;\gamma_0)$, then $\theta=\theta_0$, or, equivalently, that
\begin{equation}
\theta\neq \theta_0\Rightarrow \mathbb{M}(\theta;\gamma_0)<\mathbb{M}(\theta_0;\gamma_0). \label{key_ineq}
\end{equation}
By conditions C2 and C3, and the Arzel\`{a}--Ascoli theorem, it follows that the set $\Theta_{\epsilon}^*=\{\theta:d(\theta,\theta_0)\geq \epsilon\}$ is relatively compact, for any $\epsilon>0$, and thus its closure $\bar{\Theta}_{\epsilon}^*$ is a compact set. By conditions C1-C3 and the dominated convergence theorem it follows that the map $\theta \mapsto \mathbb{M}(\theta;\gamma_0)$ is continuous. Therefore, there exists a $\theta^*\in\bar{\Theta}_{\epsilon}^*$, such that $\sup_{\theta\in\Theta_{\epsilon}^*}\mathbb{M}(\theta;\gamma_0)=\mathbb{M}(\theta^*;\gamma_0)$, with $\theta^*\neq \theta_0$, and thus by \eqref{key_ineq} it follows that
\[
\sup_{\theta\in\Theta_{\epsilon}^*}\mathbb{M}(\theta;\gamma_0)<\mathbb{M}(\theta_0;\gamma_0),
\]
which concludes the proof of the identifiability condition 2.

For condition 3, it has been shown that for \( \phi_{0,j} \in \Phi \), there exists a corresponding \( \phi_{0,n,j} \in \Phi_{n,j} \) of order \( m \geq p + 2 \) such that \(
\|\phi_{0,n,j} - \phi_{0,j}\|_\infty \leq K q_n^{-p} = O(n^{-p\nu})
\) \citep{zhang2010spline}. This further implies \(
\|\phi_{0,n,j} - \phi_{0,j}\|_\Phi \leq K q_n^{-p} = O(n^{-p\nu})
\) \citep{zhang2010spline}. Naturally, this result applies for all \( j = 1, \dots, k \). Defining \( \theta_{0,n} = (\beta_0, \phi_{0,n}) \), if follows that:
\begin{eqnarray*}
    \mathbb{M}_n(\hat{\theta}_n; \hat{\gamma}_{n'}) - \mathbb{M}_n(\theta_0; \hat{\gamma}_{n'}) &=& \mathbb{M}_n(\hat{\theta}_n; \hat{\gamma}_{n'}) -\mathbb{M}_n(\theta_{0,n}; \hat{\gamma}_{n'}) + \mathbb{M}_n(\theta_{0,n}; \hat{\gamma}_{n'}) - \mathbb{M}_n(\theta_{0,n}; \gamma_0) \\
    && +\, \mathbb{M}_n(\theta_{0,n}; \gamma_0) - \mathbb{M}_n(\theta_{0}; \gamma_0) + \mathbb{M}_n(\theta_{0}; \gamma_0) - \mathbb{M}_n(\theta_0; \hat{\gamma}_{n'}) \\
    &=& \mathbb{M}_n(\hat{\theta}_n; \hat{\gamma}_{n'}) -\mathbb{M}_n(\theta_{0,n}; \hat{\gamma}_{n'}) + \mathbb{M}_n(\theta_{0,n}; \gamma_0) - \mathbb{M}_n(\theta_{0}; \gamma_0) + o_p(1) \\
    &\geq& \mathbb{M}_n(\theta_{0,n}; \gamma_0) - \mathbb{M}_n(\theta_0; \gamma_0) + o_p(1),
\end{eqnarray*}
where the second equality follows from the consistency of $\hat{\gamma}_{n'}$ in probability, regularity conditions C1-C3, and Taylor expansion, while the last inequality follows from the fact that $\mathbb{M}_n(\hat{\theta}_n; \hat{\gamma}_{n'}) -\mathbb{M}_n(\theta_{0,n}; \hat{\gamma}_{n'})\geq 0$ by the fact that our estimator is the maximizer of the log pseudo-likelihood over the sieve space $\Theta_n$. Now, using similar arguments to those used in \cite{zhang2010spline} and \cite{bakoyannis2017semiparametric}, leads to the conclusion that condition 3 is satisfied. This concludes the proof of Theorem~\ref{thm:consistency_p1}.

\selectlanguage{english}
\FloatBarrier
\bibliography{paper1_refs}

@article{an2009need,
  author  = {An, M. W. and Frangakis, C. E. and Musick, B. S. and Yiannoutsos, C. T.},
  year    = {2009},
  title   = {The need for double-sampling designs in survival studies: An application to monitor PEPFAR},
  journal = {Biometrics},
  volume  = {65},
  number  = {1},
  pages   = {301--306}
}

@article{bakoyannis2020semiparametric,
  author  = {Bakoyannis, G. and Zhang, Y. and Yiannoutsos, C. T.},
  year    = {2020},
  title   = {Semiparametric regression and risk prediction with competing risks data under missing cause of failure},
  journal = {Lifetime Data Analysis},
  volume  = {26},
  number  = {4},
  pages   = {659--684},
  doi     = {10.1007/s10985-020-09494-6}
}

@article{bakoyannis2017semiparametric,
  author  = {Bakoyannis, G. and Yu, M. and Yiannoutsos, C. T.},
  year    = {2017},
  title   = {Semiparametric regression on cumulative incidence function with interval-censored competing risks data},
  journal = {Statistics in Medicine},
  volume  = {36},
  number  = {23},
  pages   = {3683--3707}
}

@article{bakoyannis2012practical,
  author  = {Bakoyannis, G. and Touloumi, G.},
  year    = {2012},
  title   = {Practical methods for competing risks data: A review},
  journal = {Statistical Methods in Medical Research},
  volume  = {21},
  number  = {3},
  pages   = {257--272},
  pmid    = {21216803}
}

@article{barron1977effects,
  author  = {Barron, B. A.},
  year    = {1977},
  title   = {The effects of misclassification on the estimation of relative risk},
  journal = {Biometrics},
  volume  = {33},
  number  = {2},
  pages   = {414--418}
}

@article{brinkhof2010adjusting,
  author  = {Brinkhof, M. W. and Spycher, B. D. and Weigel, R. and Wood, R. and Messou, E. and Boulle, A. and Egger, M. and others},
  year    = {2010},
  title   = {Adjusting mortality for loss to follow-up: Analysis of five ART programmes in Sub-Saharan Africa},
  journal = {PLoS One},
  volume  = {5},
  number  = {11},
  pages   = {e14149}
}

@article{bross1954misclassification,
  author  = {Bross, I.},
  year    = {1954},
  title   = {Misclassification in 2x2 tables},
  journal = {Biometrics},
  volume  = {10},
  number  = {4},
  pages   = {478--486}
}

@article{cheng2010bootstrap,
  author  = {Cheng, G. and Huang, J. Z.},
  year    = {2010},
  title   = {Bootstrap consistency for general semiparametric M-estimation},
  journal = {The Annals of Statistics},
  volume  = {38},
  pages   = {2884--2915}
}

@article{edwards2013accounting,
  author  = {Edwards, J. K. and Cole, S. R. and Troester, M. A. and Richardson, D. B.},
  year    = {2013},
  title   = {Accounting for misclassified outcomes in binary regression models using multiple imputation with internal validation data},
  journal = {American Journal of Epidemiology},
  volume  = {177},
  number  = {9},
  pages   = {904--912}
}

@article{egger2011correcting,
  author  = {Egger, M. and Spycher, B. D. and Sidle, J. and Weigel, R. and Geng, E. H. and Fox, M. P. and MacPhail, P. and others},
  year    = {2011},
  title   = {Correcting mortality for loss to follow-up: A nomogram applied to antiretroviral treatment programmes in Sub-Saharan Africa},
  journal = {PLoS Medicine},
  volume  = {8},
  number  = {1}
}

@article{geng2008sampling,
  author  = {Geng, E. H. and Emenyonu, N. and Bwana, M. B. and Glidden, D. V. and Martin, J. N.},
  year    = {2008},
  title   = {Sampling-based approach to determining outcomes of patients lost to follow-up in antiretroviral therapy scale-up programs in Africa},
  journal = {JAMA},
  volume  = {300},
  number  = {5},
  pages   = {506--507}
}

@article{greenland1988variance,
  author  = {Greenland, S.},
  year    = {1988},
  title   = {Variance estimation for epidemiologic effect estimates under misclassification},
  journal = {Statistics in Medicine},
  volume  = {7},
  number  = {7},
  pages   = {745--757}
}

@article{ha2015semiparametric,
  author  = {Ha, J. and Tsodikov, A.},
  year    = {2015},
  title   = {Semiparametric estimation in the proportional hazard model accounting for a misclassified cause of failure},
  journal = {Biometrics},
  volume  = {71},
  number  = {4},
  pages   = {941--949},
  doi     = {10.1111/biom.12338}
}

@book{kalbfleisch2002statistical,
  author    = {Kalbfleisch, J. D. and Prentice, R. L.},
  year      = {2002},
  title     = {The Statistical Analysis of Failure Time Data},
  edition   = {2},
  publisher = {John Wiley and Sons},
  address   = {New York},
  doi       = {10.1002/9781118032985}
}

@book{kosorok2008introduction,
  author    = {Kosorok, M. R.},
  year      = {2008},
  title     = {Introduction to Empirical Processes and Semiparametric Inference},
  publisher = {Springer},
  address   = {New York}
}

@article{lyles2011validation,
  author  = {Lyles, R. H. and Tang, L. and Superak, H. M. and King, C. C. and Celentano, D. D. and Lo, Y. and Sobel, J. D.},
  year    = {2011},
  title   = {Validation data-based adjustments for outcome misclassification in logistic regression: An illustration},
  journal = {Epidemiology},
  volume  = {22},
  number  = {4},
  pages   = {589--597}
}

@article{magder1997logistic,
  author  = {Magder, L. S. and Hughes, J. P.},
  year    = {1997},
  title   = {Logistic regression when the outcome is measured with uncertainty},
  journal = {American Journal of Epidemiology},
  volume  = {146},
  number  = {2},
  pages   = {195--203}
}

@article{monroy2024alcohol,
  author  = {Monroy, A. and Goodrich, S. and Brown, S. A. and et al.},
  year    = {2024},
  title   = {Effects of Alcohol Use on Patient Retention in HIV Care in East Africa},
  journal = {AIDS and Behavior},
  volume  = {28},
  pages   = {4020--4028},
  doi     = {10.1007/s10461-024-04483-z}
}

@article{mpofu2020pseudo,
  author  = {Mpofu, P. B. and Bakoyannis, G. and Yiannoutsos, C. T. and Mwangi, A. W. and Mburu, M.},
  year    = {2020},
  title   = {A pseudo-likelihood method for estimating misclassification probabilities in competing-risks settings when true-event data are partially observed},
  journal = {Biometrical Journal},
  volume  = {62},
  number  = {7},
  pages   = {1747--1768},
  doi     = {10.1002/bimj.201900198}
}

@article{neuhaus1999bias,
  author  = {Neuhaus, J. M.},
  year    = {1999},
  title   = {Bias and efficiency loss due to misclassified responses in binary regression},
  journal = {Biometrika},
  volume  = {86},
  number  = {4},
  pages   = {843--855}
}

@article{robins2000inference,
  author  = {Robins, J. M. and Wang, N.},
  year    = {2000},
  title   = {Inference for imputation estimators},
  journal = {Biometrika},
  volume  = {87},
  number  = {1},
  pages   = {113--124}
}

@article{shen1994convergence,
  author  = {Shen, X. and Wong, W. H.},
  year    = {1994},
  title   = {Convergence rate of sieve estimates},
  journal = {The Annals of Statistics},
  volume  = {22},
  pages   = {580--615}
}

@article{spiegelman2001efficient,
  author  = {Spiegelman, D. and Carroll, R. J. and Kipnis, V.},
  year    = {2001},
  title   = {Efficient regression calibration for logistic regression in main study/internal validation study designs with an imperfect reference instrument},
  journal = {Statistics in Medicine},
  volume  = {20},
  number  = {1},
  pages   = {139--160}
}

@article{tang2015binary,
  author  = {Tang, L. and Lyles, R. H. and King, C. C. and Celentano, D. D. and Lo, Y.},
  year    = {2015},
  title   = {Binary regression with differentially misclassified response and exposure variables},
  journal = {Statistics in Medicine},
  volume  = {34},
  number  = {9},
  pages   = {1605--1620}
}

@article{tenenbein1970double,
  author  = {Tenenbein, A.},
  year    = {1970},
  title   = {A double sampling scheme for estimating from binomial data with misclassifications},
  journal = {Journal of the American Statistical Association},
  volume  = {65},
  number  = {331},
  pages   = {1350--1361}
}

@book{vanderVaart1996weak,
  author    = {van der Vaart, A. W. and Wellner, J. A.},
  year      = {1996},
  title     = {Weak Convergence and Empirical Processes with Applications to Statistics},
  publisher = {Springer-Verlag},
  address   = {New York}
}

@article{yiannoutsos2008sampling,
  author  = {Yiannoutsos, C. T. and An, M. W. and Frangakis, C. E. and Musick, B. S. and Braitstein, P. and Wools-Kaloustian, K. and Ochieng, D.},
  year    = {2008},
  title   = {Sampling-based approaches to improve estimation of mortality among patient dropouts: Experience from a large PEPFAR-funded program in Western Kenya},
  journal = {PLoS One},
  volume  = {3},
  number  = {12},
  pages   = {e3843}
}

@article{zhang2010spline,
  author  = {Zhang, Y. and Hua, L. and Huang, J.},
  year    = {2010},
  title   = {A spline-based semiparametric maximum likelihood estimation method for the Cox model with interval-censored data},
  journal = {Scandinavian Journal of Statistics},
  volume  = {37},
  pages   = {338--354}
}

@article{rubin1996multiple,
  title={Multiple imputation after 18+ years},
  author={Rubin, Donald B},
  journal={Journal of the American statistical Association},
  volume={91},
  number={434},
  pages={473--489},
  year={1996},
  publisher={Taylor \& Francis}
}

\end{document}